\documentclass{emulateapj}

\usepackage{color}
\usepackage{lscape}

\newcommand{\simgt}{\lower.5ex\hbox{$\; \buildrel > \over \sim \;$}}
\newcommand{\simlt}{\lower.5ex\hbox{$\; \buildrel < \over \sim \;$}}

\def\btheta{\mbox{\boldmath $\theta$}}

\newcommand{\Mvir}{M_{\rm vir}}
\newcommand{\cvir}{c_{\rm vir}}
\newcommand{\Msun}{M_{\odot}}

\slugcomment{}

\shorttitle{Subaru weak-lensing study of A2163: bimodal mass structure}
\shortauthors{Okabe et al.}

\begin{document}

\title{ Subaru weak-lensing study of A2163: bimodal mass structure
   \altaffilmark{*}}
   \altaffiltext{*}{This work is based in part on data collected at Subaru Telescope and obtained from
           the SMOKA, which is operated by the Astronomy Data Center, National Astronomical Observatory of Japan.}



\author{N. Okabe\altaffilmark{1},
H. Bourdin\altaffilmark{2},
P. Mazzotta\altaffilmark{2,3},
AND
S. Maurogordato\altaffilmark{4}.
}


\altaffiltext{1}{Academia Sinica Institute of Astronomy and Astrophysics (ASIAA), P.O. Box 23-141, Taipei 10617, Taiwan;okabe@asiaa.sinica.edu.tw}

\altaffiltext{2}{Dipartimento di Fisica, Universit\`a degli Studi di
Roma `Tor Vergata', via della Ricerca Scientifica 1, 00133 Roma,
Italy} 

\altaffiltext{3}{Harvard-Smithsonian Center for Astrophysics, 60 Garden Street,
Cambridge, MA 02138, USA}

\altaffiltext{4}{Universit\'e de Nice Sophia-Antipolis, 
CNRS, Laboratoire Cassiop\'ee, CNRS, UMR 6202, Observatoire
de la C\^ote d' Azur, BP4229, 06304 Nice Cedex 4, France}

\begin{abstract}
We present a weak-lensing
analysis of the merging cluster A2163 using Subaru/Suprime-Cam and CFHT/Mega-Cam data and
discuss the dynamics of this cluster merger, based on complementary
weak-lensing, X-ray, and optical spectroscopic data sets. From two-dimensional 
multi-component weak-lensing analysis, we reveal that the
cluster mass distribution is well described by three main components including
a two component main cluster A2163-A with mass ratio 1:8, and its
cluster satellite A2163-B. The bimodal mass distribution in
A2163-A is similar to the galaxy density distribution, but appears as
spatially segregated from the brightest X-ray emitting gas region.
We discuss the possible origins of this gas-dark matter
offset, and suggest the gas core of the A2163-A subcluster has been
stripped away by ram pressure from its dark matter component. The
survival of this gas core to the tidal forces exerted by the main
cluster lets us infer a subcluster accretion with a non-zero impact
parameter. Dominated by the most massive component of A2163-A, the mass
distribution of A2163 is well described by a universal Navarro-Frenk-White profile as
shown by a one-dimensional tangential shear analysis, while the
singular-isothermal sphere profile is strongly ruled out.
Comparing this cluster mass profile with profiles derived assuming intracluster medium
hydrostatic equilibrium (H.E.) in two opposite regions of the cluster
atmosphere has allowed us to confirm the prediction of a departure
from H.E. in the eastern cluster side, presumably due to shock heating. 
Yielding a cluster mass estimate of $M_{500}=11.18_{-1.46}^{+1.64}\times10^{14}h^{-1}\Msun$, 
our mass profile confirms the exceptionally high mass of A2163,
consistent with previous analyses relying on the cluster dynamical analysis and
$Y_{\rm X}$ mass proxy.
\end{abstract}





\section{Introduction}
Galaxy clusters are the largest self-gravitating systems in the
universe, residing at the intersection of large-scale filamentary structures of the cosmic web. 
According to the hierarchical structure formation scenario based on a cold dark matter (CDM) paradigm,
mass accretion flows onto clusters are still ongoing and high-mass
ratio cluster mergers, the so-called major mergers, sometimes occur.
Major cluster mergers are among the most energetic events in
the universe, releasing amounts of gravitational energy as large as 
$10^{64-65} {\rm erg}$. Spatially resolved X-ray observation has revealed to us
that this collision energy partly dissipates in the intracluster medium (ICM) through 
shock heating and turbulence, yielding a complex ICM brightness and thermal structure 
\citep[see, e.g.,][and reference therein]{mar07}. The dynamics of cluster collisions are however 
dominated by the cluster mass distribution, which cannot be constrained from X-ray observations alone. 
Due to the collisional nature of ICM, cluster mergers are indeed expected to generate some transient 
decoupling between spatial distributions of the CDM and hot X-ray emitting gas. Moreover, 
mass estimates relying on gas properties may be biased by merger-induced perturbations of the ICM hydrostatic 
equilibrium (H.E.) within individual colliding clusters.

Weak-lensing distortions of background galaxy images provide us with a
unique opportunity to reconstruct the distribution of matter in clusters 
without any assumption of mass model and dynamical states
\citep[e.g., ][]{kai95, bar01, sch06, oka08, oka10b} 
and measure cluster masses
\citep[e.g., ][]{gra02, gav03, bar07, hoe07, oka10b, ume09}.
Therefore, a weak-lensing analysis of merging clusters \citep[e.g.,][]{clo06,mah07,oka08,oka10b,mer11}
is an observational breakthrough in measuring mass distribution, thereby providing
complementary information to X-ray measurements.  
\cite{oka08} conducted a systematic study of seven merging clusters, 
representing various merging stages and conditions, based
on a joint weak-lensing, optical photometric, and X-ray analysis,
and revealed that the mass and optical light of member galaxies are similarly
distributed in merging clusters regardless of their merging stages, but 
the mass distribution in merging clusters is highly irregular, and is
quite different from the ICM distributions. It indicates that the gas
and mass evolutions under cluster mergers are different.
This feature is confirmed by weak-lensing studies of 30 clusters as a collaboration of ``Local 
Cluster Substructure Survey'' \citep{oka10b}.
Thus, a joint analysis of clusters
\citep[e.g.,][]{mah08,oka08,oka10c,kaw10,ume10,zha10} 
will yield a comprehensive and quantitative understanding of cluster merger physics
involved in the structure formation. 

Located at a redshift of $0.203$,  A2163 is a rich, X-ray luminous and hot galaxy cluster showing 
various signatures of ongoing merger events, 
including irregular optical and X-ray morphologies \citep[see, e.g.][]{elb95,mar01,mau08,bou10},
 and a prominent radio halo emission \citep{fer01,fer04}. This cluster has been 
known to exhibit a clear spatial segregation between a bimodal galaxy distribution and a more centrally peaked ICM morphology \citep[][hereafter M08]{mau08}, 
while the projected distance separating its main mass and gas centroids has been measured as one of the largest in a sample of 38 clusters analyzed from both strong galaxy lensing and X-ray imaging \citep[][]{shan10}. 
As revealed from spectroscopic and photometric analyses in M08, 
the galaxy distribution in A2163 can be separated into two components: 
the massive cluster A2163-A and its northern companion A2163-B, and 
the main cluster component itself showing a bimodal morphology with two brightest galaxies (BCG1 and BCG2). 
In a more recent analysis of \textit{Chandra} and \textit{XMM-Newton} data, \citet[][hereafter B11]{bou10} evidenced the westward motion of a cool core across the E-W elongated  atmosphere of the main cluster, A2163-A. 
Located close to the second galaxy over-density, 
this gas 'bullet' appears to have been spatially separated from its galaxy component as a result of high-velocity accretion. 
From gas brightness and temperature profile analysis performed in two opposite regions of the main cluster, 
B11 further showed that the ICM has been adiabatically compressed behind this crossing ``bullet''.

Characterized by an exceptionally high ICM temperature 
first measured in X-ray \citep[$k_BT=13.9$~keV;][]{arn92} 
then confirmed from its Sunyaev-Zeldovich (SZ) distortion \citep{Nord09}, 
A2163 has been suggested to be exceptionally 
massive from various X-ray and SZ analyses \citep[see, e.g.][]{vik09,Planck11}
via the mass scaling relation.
Former weak-lensing analyses have been conducted using 
the Very Large Telescope and CFHT/Mega-Cam analyses of \citet[][]{cyp04} and \citet[][]{rad08}.
These analyses, however, did not exclude merger galaxies in the background shear catalog, 
as demonstrated by other studies \citep{bro05,oka08,ume08,ume09,oka10b,ume10}.
As shown by \cite{oka10b}, 
a contamination of member galaxies significantly dilutes lensing distortion signals, 
mainly in the central region, thereby yielding biased estimations on cluster parameters.

In this paper, we conducted a weak-lensing analysis using Subaru/Suprime-Cam
and made a secure selection to avoid a contamination of member
galaxies in the shear catalog, combining CFHT/Mega-Cam data.
The structure of this paper is as follows:  in Section~\ref{sec:data} we
briefly describe the weak-lensing analysis. The projected
distributions of mass, member galaxies and hot-gas are compared in
Section \ref{sec:map}. Section \ref{sec:1d} is devoted to a one-dimensional 
shear analysis to measure cluster mass.
In Sec \ref{sec:2d}, 
we perform, based on a two dimensional shear pattern, 
a multi-components fitting to measure three components revealed by M08. 
We compare weak lensing mass with 
dynamical and X-ray hydrostatic equilibrium masses in Section ~\ref{sec:mass}.
In Section ~\ref {sec:dis}, we discuss
quantitatively the physical process of making an offset between mass and gas distributions.
Section ~\ref{sec:sum} summarizes our results.
The cosmological parameters of $H_0 = 100h^{-1}$ Mpc$^{-1}$ km s$^{-1}$,
$\Omega_0=0.3$, and $\Lambda_0=0.7$ are used in this paper. Given the
cosmology, $1\farcm=140.21~h^{-1}{\rm kpc}$.

\section{Data Analysis} \label{sec:data}

We retrieved $R_{\rm c}$ image data from the Subaru archival data
(SMOKA\footnote{http://smoka.nao.ac.jp/index.jsp}). 
The data were reduced by the standard imaging process of using the
reduction software for Suprime-Cam, SDFRED \citep{ouc04}, as described in \citet{oka08} and \citet{oka10b}.
Astrometry calibration was conducted by fitting the final stacked image with the Two Micron All Sky Survey data catalog.
The residual in astrometric fitting was lower than the CCD pixel size.
The exposure time is $75$ minutes and the seeing is $0\farcs79$;
therefore, it is suitable for our weak-lensing study.
Since no other band data are available, we use CFHT/Mega-cam $g'$
imaging data to exclude unlensed galaxies in the shear source catalog via a
color-magnitude plane.

\begin{figure*}
  \begin{center}
   \includegraphics[width=0.55 \textwidth,angle=0,clip]{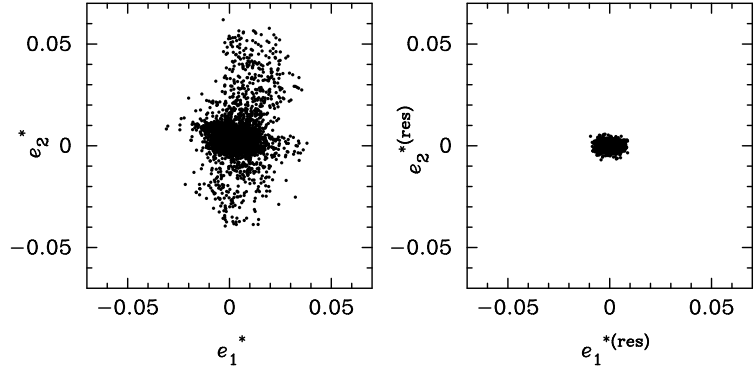} 
  \end{center}
  \caption{Stellar ellipticity distributions before and after the PSF anisotropy
 correction.  The left panel shows the raw ellipticity
 components $(e^*_1,e^*_2)$ of stellar objects, 
and the right panel shows the residual ellipticity components 
$(e^{{\rm res,*}}_1, e^{{\rm res,*}}_2)$ after the PSF anisotropy correction.}
\label{fig:stell_e}
\end{figure*}

\begin{figure*}
  \begin{center}
   \includegraphics[width=0.55 \textwidth,angle=0,clip]{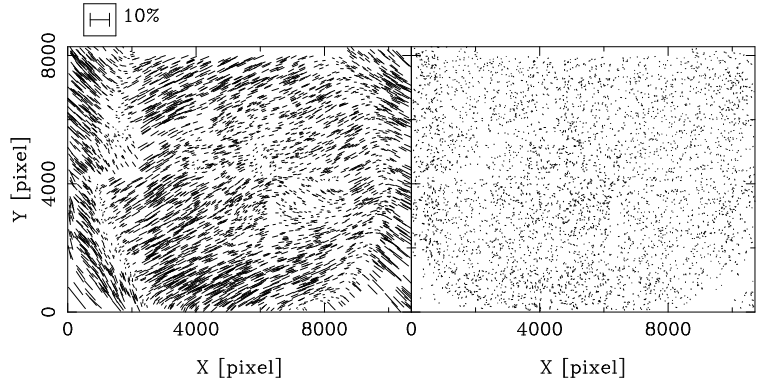} 
  \end{center}
  \caption{ Distortion field of stellar ellipticities before (left) and after (right) the PSF
 anisotropy correction. 
The orientation of the sticks indicates the position angle of the major
 axis of stellar ellipticity, whereas
the length is proportional to the modulus of stellar ellipticity.
A stick with the length of $10\%$ ellipticity is shown above the left panel.
}
\label{fig:emap}
\end{figure*}

\begin{figure*}
  \begin{center}
   \includegraphics[width=0.55 \textwidth,angle=0,clip]{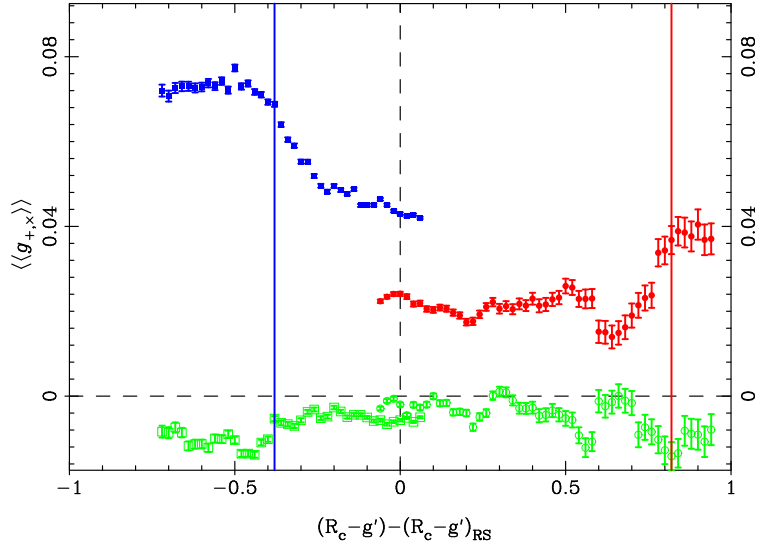} 
  \end{center}
  \caption{ Mean distortion strength $\langle\langle g_{+,\times}\rangle\rangle$
     over the radii of
    $1'\le r\le 18'$, as a function of color. The background samples are defined with
    galaxies redder or bluer than the red-sequence by at least the
    color offsets given by the $x$-label. The distortion strength is
    changed due to dilution by cluster members. The red circle and blue square points denote
    the mean tangential distortion $\langle\langle g_{+}\rangle\rangle$ for redder and 
    bluer background samples, respectively. 
    The green circle and square points denote the mean tangential distortion for 
    the $45$ degree rotated component $\langle\langle g_{\times}\rangle\rangle$ for redder and 
    bluer samples, respectively.
    The two vertical solid lines denote our choices of the color cuts used to
    define the red/blue background galaxy samples shown in
    Figure~\ref{fig:color-mag}. The details are described in \cite{ume08} and \cite{oka10b}.
   }
\label{fig:dilution}
\end{figure*}

\begin{figure*}
  \begin{center}
   \includegraphics[width=0.35 \textwidth,angle=-90,clip]{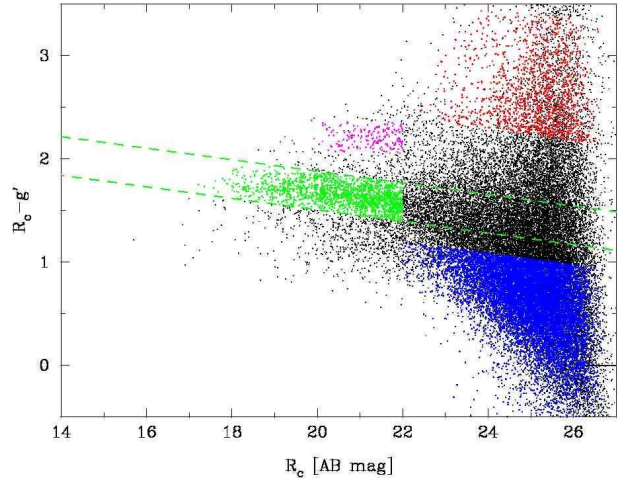} 
  \end{center}
  \caption{Color-magnitude diagram. 
Green points are member galaxies of which spatial distribution is shown in the top right
and bottom left panels of Figure \ref{fig:kappa+den}.
Two green dashed lines represent the width of the red sequence.
The red and blue points are the
background galaxy samples, used for the lensing distortion analysis,
whose colors are redder and blue than the red sequence, respectively. 
The magenta points denote a sample of galaxies in the right bottom panel of Figure \ref{fig:kappa+den}.
}
\label{fig:color-mag}
\end{figure*}

Our weak-lensing analysis is done using the IMCAT package provided by
N. Kaiser (Kaiser et al. 1995\footnote{http://www.ifa.hawaii/kaiser/IMCAT}).  
We use the same pipeline as \citet{oka10a,oka10b} with some
modifications followed by \citet{erb01}. 
We first measure the image ellipticity, $e_\alpha$, from the
weighted quadrupole moments of the surface brightness of each object
and then correct the point-spread function (PSF) anisotropy by solving $e_\alpha'=e_\alpha-P_{\rm
sm}^{\alpha\beta}(P_{\rm sm}^*)^{-1}_{\beta\gamma}e^{\gamma*}$, where $P_{\alpha\beta}$ is the
smear polarizability tensor and the asterisk denotes the stellar objects.
We fit the stellar anisotropy kernel $(P_{\rm sm}^*)^{-1}_{\alpha\beta}e^{\beta*}$ with the
second-order bi-polynomials function in several subimages whose sizes
are determined based on the typical coherent scale of the measured PSF
anisotropy pattern. The median stellar ellipticities before and after the anisotropic
correction are $(\bar{e}_1^*, \bar{e}_2^*)=(1.45\pm0.02, 1.52\pm0.01)\times10^{-2}$
and  $(\bar{e}_1^{\rm res,*}, \bar{e}_2^{\rm
res,*})=(-3.74\pm5.30,-5.07\pm4.73)\times10^{-5}$, respectively.
Stellar ellipticities before and after the correction and their pattern are shown in
Figures \ref{fig:stell_e} and \ref{fig:emap}, respectively.
We next estimate the reduced shear $g_\alpha=\gamma_{\alpha}
/(1-\kappa)=(P_g)_{\alpha\beta}^{-1}e_{\beta}'$ using the pre-seeing shear
polarizability tensor $P_g$, where we adopt the scalar value
$(P_g)_{\alpha\beta}={\rm Tr}[P_g]\delta_{\alpha\beta}/2$. 
We select background galaxies in the range of ${\bar
r}_h^*+\sigma_{r_h^*}<r_h<r_{h,\rm max}\simeq6.6~{\rm pixels}$,  
where $r_h$ is the half-light radius, and ${\bar r}_h^*$ and
$\sigma_{r_h^*}$ are the median and standard error of
stellar half-light radii, $r_{h^*}$, corresponding to the half median
width of circularized PSF. 
Then, we make a secure selection of background galaxies in a
color-magnitude plane in order to minimize a dilution of the
weak-lensing signals caused by a contamination of unlensed galaxies,
mainly member galaxies.
As shown in \cite{oka10b}, the dilution effect on the lensing signals is
more pronounced at smaller radii because the ratio of the number density
of cluster galaxies to background ones rises towards the inner region.
We use Subaru/Suprime-Cam $R_{\rm c}$ and CFHT/Mega-cam $g'$ images. 
The difference in filter sensitivity functions of the wavelength in the
$g'$ band between these two instruments is negligible.
Therefore, a combination with
Subaru and CFHT enables us to efficiently exclude member galaxies inthe  background
shear catalog. 
Following \cite{ume08}, we calculate the lensing signal as a function of
color (Figure \ref{fig:dilution}), by averaging the tangential distortion strengths \citep[see
also][]{oka10b}. We securely select background galaxies in the range of
$0.82<(R_{\rm c}-g')-(R_{\rm c}-g')_{{\rm RS}}<2$, $-2<(R_{\rm
c}-g')-(R_{\rm c}-g')_{{\rm RS}}<-0.38$ and $22~{\rm ABmag}<R_{\rm c}$, where $(R_{\rm c}-g')_{{\rm
RS}}$ is the best-fit linear function as a magnitude $R_{\rm c}$ of
red-sequence galaxies; $(R_{\rm c}-g')_{{\rm RS}}=-0.056R_{\rm c}+2.812$,
as shown in Figure \ref{fig:color-mag}.
Here, the magnitudes and colors are measured by MAG\_AUTO (i.e., total
magnitude) and MAG\_APER (i.e., aperture magnitude) in SExtractor, respectively.
By a selection in the color magnitude plane, 
the number density of background galaxies in the shear catalog decreases
from $\simeq23$ to $\simeq12~\rm arcmin^{-2}$. 
The mean redshift of these background galaxies is estimated by matching
the COSMOS catalog \citep{ilb09} to calculate the average lensing
weight $\langle D_{\rm ls}/D_{\rm s}\rangle=\int_{z_d}dz d P_{\rm
WL}/dz D_{\rm ls}/D_{\rm s}$, where $D_{\rm s}$ and $D_{\rm ls}$ are the angular diameter
distances between the observer and source (background galaxy) and lens
and source, respectively, and $dP_{\rm WL}/dz$ is a probability function
of redshift distribution. We obtain $\langle D_{\rm ls}/D_{\rm s}\rangle\simeq0.7209$.

\section{Maps of Mass, Galaxies and Gas}\label{sec:map}


According to the CDM paradigm, the collisional ICM is expected to experience 
a spatial decoupling from galaxies and dark matter during cluster collisions. 
Comparing hot X-ray emitting gas maps with the cluster member galaxy density and mass distributions 
has thus been known to testify to the merging cluster dynamical states \citep[see, e.g.][]{clo06,oka08}.
In order to investigate the collision scenario in A2163, we map out the cluster mass distribution and 
compare it with spatial distributions of the galaxy and hot gas components.

As described in detail in \citet{oka08}, we pixelize the shear
pattern into a regular grid using Gaussian smoothing kernel 
$w_g(\theta)\propto \exp[-\theta^2/\theta_g^2]$ with $\theta_g = {\rm
FWHM}/\sqrt{4\ln{2}}$. 
We adopt the smoothing scale of ${\rm FWHM}=1\farcm33$ due to the limitation
of the number of background galaxies. We also use a statistical weight 
for each background galaxy in the context of 
\begin{eqnarray}
 u_{g,i}=\frac{1}{\sigma_{g,i}^2+\alpha^2}, \label{eq:ug}
\end{eqnarray}
where $\sigma_{g,i}$ is the rms error for shear measurement of $i$th
galaxy and $\alpha$ is the softening constant variance. We set
$\alpha=(\sum \sigma_{g,i}^2/N)^{1/2}\simeq0.44$.
The reduced shear at pixel position of $\btheta_n$ is obtained by
\begin{equation}
\label{eq:smshear}
\langle{g_{\alpha}}\rangle(\btheta_n) = \frac{\sum_i w_g(\btheta_n-\btheta_i) u_{g,i} g_{\alpha,i}}{\sum_i
w_g(\btheta_n-\btheta_i) u_{g,i}}.
\end{equation}
Then, we invert the pixelized reduced shear (Equation \ref{eq:smshear}) with a
weight of the inverse of the variance at each pixel to the lensing
convergence field, based on the Kaiser \& Squires inversion method \citep{kai93}. 

We also map out distributions of optical luminosities and the number density of member
galaxies with the same kernel with $u_g=1$. 
The luminosity and density maps are sensitive to luminous and high-density structures, respectively. 
We selected red-sequence galaxies with $R_{\rm c}<22{\rm ABmag}$ as
member galaxies. The absolute magnitudes for member galaxies are
calculated from the apparent magnitude using the $k$-correction for early-type
galaxies. We assume that all cluster member galaxies in the catalog are at a single redshift.

\begin{figure*}
  \begin{center}
   \includegraphics[width=0.55 \textwidth,angle=0,clip]{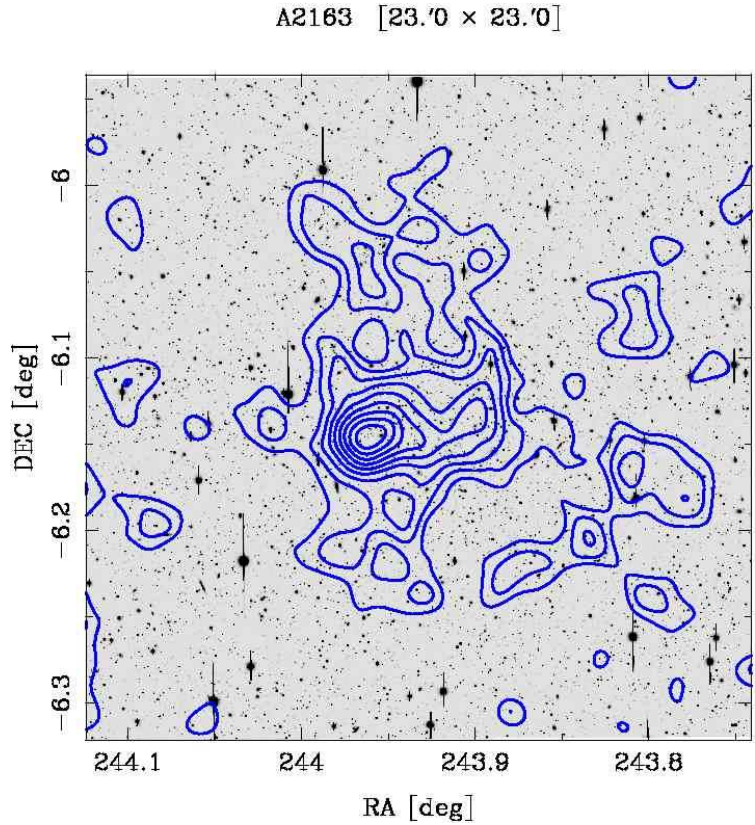} 
  \end{center}
  \caption{Subaru $R_{\rm c}$-band image ($23\farcm \times 23\farcm$)
 overlaid with contours of the $\kappa$-field reconstructed from
the weak-shear field, in units of $1\sigma$ reconstruction error
 $\delta\kappa=0.0404$.
}
\label{fig:opt+kappa}
\end{figure*}
\begin{figure*}
  \begin{center}
   \includegraphics[width=0.85 \textwidth,angle=0,clip]{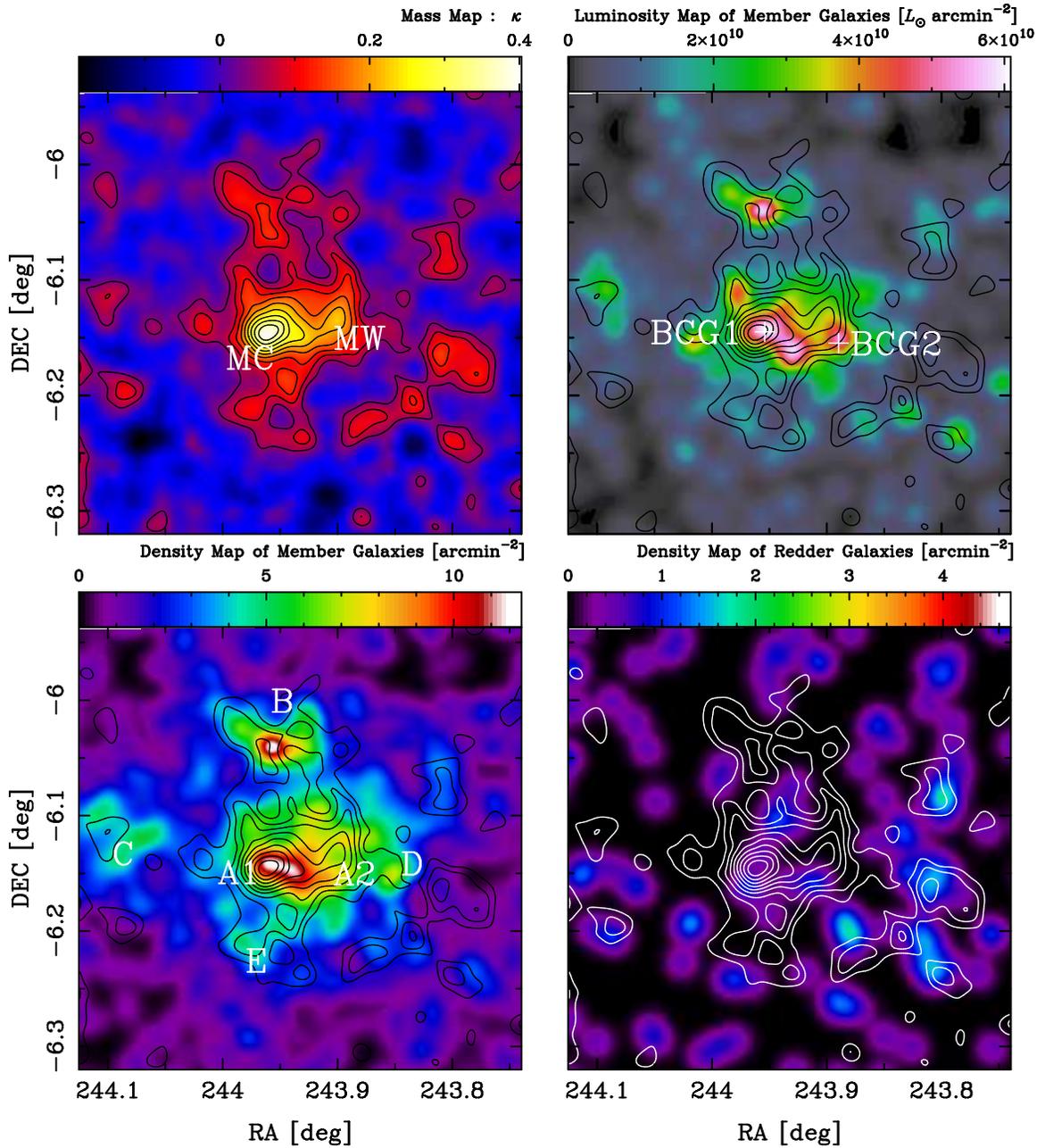} 
  \end{center}
  \caption{Top left: reconstructed lensing $\kappa$-field in 
 units of $1\sigma$ reconstruction error. The bimodal structure in the
 central region is clearly found.  The significance levels of bimodal
 peaks are
 $\sim10\sigma$ and $\sim5.6\sigma$, which are referred to as MC and MW,
 respectively. 
Top right : luminosity map for red-sequence galaxies in the $R_{\rm c}$ band. Overlaid are the
 contours of the $\kappa$ field.  Two crosses denote BCG positions.
Bottom left : density map for red-sequence galaxies in units of arcmin$^{-2}$. The four density
 clumps other than the bimodal structure, discovered by M08 \citep{mau08}, are confirmed. 
All labels for the density clump are quoted from M08.
Bottom right : density map for galaxies whose color is redder by
 $\sim0.6$ at $R_{\rm c}=22~{\rm ABmag}$ than those of member galaxies.
}
\label{fig:kappa+den}
\end{figure*}

The resultant lensing $\kappa$ field is shown in Figures
\ref{fig:opt+kappa} and \ref{fig:kappa+den}, with contours equi-spaced in
units of $1\sigma$ reconstruction error, $\delta\kappa=0.0404$, above
the $1\sigma$ level. This mass distribution clearly exhibit a bimodal morphology 
in the central region ($r\sim3\farcm$), with two peaks reaching significance levels
of $\sim10\sigma$ and $\sim5.6\sigma$, respectively. 
Interestingly, these two peaks 
coincide with the two galaxy overdensities A1 and A2 hosting the two brightest cluster galaxies 
revealed in M08, BCG1, and BCG2 (see also the top right and bottom left panels of Figure
\ref{fig:kappa+den}). We hereafter refer to these primary and secondary mass 
peaks as MC and MW, respectively. 
The cluster mass distribution further exhibits some anisotropies at larger radii, 
mostly coinciding with peripheral optical clumps first revealed in M08 and 
presently confirmed (see the bottom left panel of Figure \ref{fig:kappa+den}).
The most significant of these coinciding mass and optical peaks is $\sim3.3\sigma$ at the optical clump, B. 

The mass distribution in the region of optical substructures C, D, and E, discovered by M08,
is similar to those of the number and luminous density, albeit low significance levels (Table \ref{tab:opticalclumps}).
We measure luminosities within $2\farcm$ centering each $\kappa$ peak (Table \ref{tab:opticalclumps}), 
where a background region of $20-22\farcm$ centering BCG1 is used. 
We find that luminosities and signal-to-noise ratios of $\kappa$ are correlated.
We tried to measure model-independent mass for each optical subclump, following \cite{oka10a}.
However, since the result is sensitive to the choice of background region, we could not obtain reliable results.
An over-density region is further revealed from the density of galaxies whose colors are redder by $\sim0.6$ at $R_{\rm c}=22~{\rm ABmag}$ (see the bottom right panel of Figure \ref{fig:kappa+den}). 
The width of colors for these galaxies is $\sim0.15$. 
Distributed at the west of A2163, these optical structures are associated with mass clumps of $\sim2-3\sigma$. 
More generally speaking, 
the cluster mass distribution appears as spatially correlated with density distribution of its member galaxies, 
consistently with weak-lensing studies in other clusters \citep{oka08,oka10b}.

\begin{figure*}
  \begin{center}
   \includegraphics[width= \textwidth,angle=0,clip]{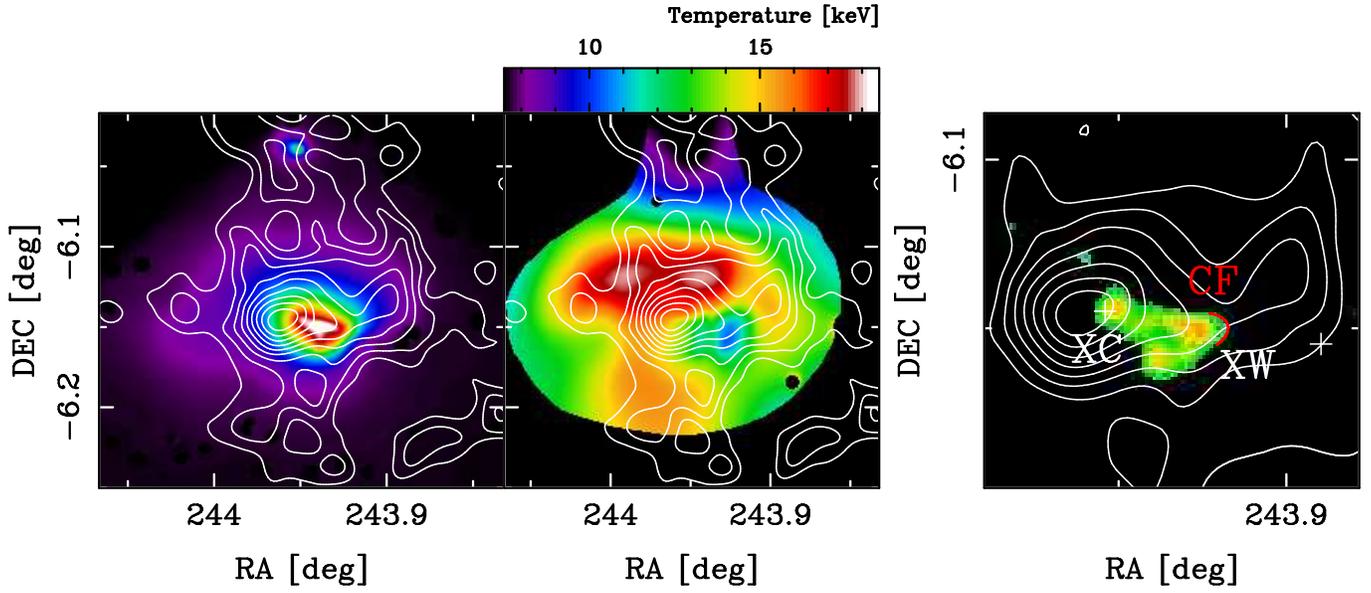}
  \end{center}
  \caption{ Left: adaptively-smoothed {\it Chandra} X-ray image
 ($14\farcm\times14\farcm$) in the
 $0.5-2.5$ keV band. Overlaid are the contours of the lensing $\kappa$-field in
 units of $1\sigma$ reconstruction error (Figure \ref{fig:kappa+den}). Middle : temperature map from {\it
 XMM-Newton} data with the same mass contours as those in the left panel. 
Right: high resolution residual {\it Chandra} image obtained after
 subtracting large scales of the denoised wavelet transform of the X-ray
 image. Two components (XC and XW) are clearly found in the X-ray
 emitting core. Two crosses denote BCG positions. 
Red (color version) or dark gray (black and white version) curved line
represents the cold front (CF; see B11).
Overlaid are the same mass contours above the $3\sigma$
significance. The offsets between the X-ray core and two mass peaks are clearly found.
}
\label{fig:chan+kappa}
\end{figure*}

We next compare the cluster mass contours with X-ray surface brightness and temperature maps
derived from \textit{XMM-Newton} and \textit{Chandra data} analysis in B11. 
The left panel of Figure
\ref{fig:chan+kappa} shows us a wavelet denoised map of the overall cluster atmosphere, with mass 
contours superimposed. 
These two distributions clearly exhibit a spatial segregation, the X-ray emission 
appearing as unimodal and E-W elongated, with an emission peak located between the two mass peaks. 
The right panel of Figure \ref{fig:chan+kappa} shows high-resolution details in a multi-resolution analysis 
of the {\it Chandra} image. 
This analysis reveals a secondary cluster X-ray core, XW, separated from the main 
cluster emission peak, XC. 
As shown in B11, this wedge-shaped feature (red-curved line in the right panel of Figure \ref{fig:chan+kappa}) 
appears as a stripped cool core embedded
in the hotter ICM of A2163-A and delimited in the westward direction by a cold front.
The evidence of this cold front
let us infer the westward motion of this gas 'bullet' along the elongated atmosphere of A2163-A. 
Interestingly, we
now observe that the crossing core appears as preceded by the secondary mass peak, 
from which it might have 
been spatially separated during a subcluster accretion onto the main cluster. 
This offset feature between the X-ray clump and mass peak is consistent with other cold front clusters on a merging phase \citep{clo06,oka08,oka10b}. 
It suggests a possibility of ram pressure stripping of the accreted cluster core (see also Section \ref{sec:dis}).

It is further worth noticing that the main cluster mass peak, MC, also appears 
as slightly offset from the X-ray core, XC, so that the centroid of the gas core of XC and XW is at the intermediate position between the two mass peaks. 
This `double offset' separating the gas and dark matter contents of both the infalling and main cluster cores 
presents some similarities to the fast accretion observable in 1E0657-56 \citep{clo06}, 
the so-called ``bullet-cluster''. 
To our knowledge, no other bimodal merger has been known so far to exhibit such a significant 
offset between the gas and dark matter components of its two components \citep[see, e.g. other examples in ][]{oka08}.



The northern subcluster A2163-B exhibits a spatial coincidence between X-ray, 
mass, and member galaxies, consistent with what is usually observed in 
pre-merging systems \citep{oka08}. Consistent with the lack 
of any interaction evidence found from ICM thermodynamics 
between A2163-A and A2163-B, this spatial coincidence 
suggests that A2163-B is likely to be observed before interacting with A2163-A. 

\section{Tangential distortion Analysis} \label{sec:1d}

We conduct a tangential distortion study, which is
a one-dimensional lensing analysis, in order to measure the total cluster mass. 
The tangential distortion component of the reduced shear,
$g_\alpha=(g_1,g_2)$, and the $45$ degree
rotated component for individual galaxies ($i$th galaxy) are obtained by
\begin{eqnarray}
g_{+,i}&=&-g_{1,i}\cos2\varphi-g_{2, i}\sin2\varphi, \nonumber\\
g_{\times,i}&=&-g_{1,i}\sin2\varphi+g_{2,i}\cos2\varphi, 
\label{eq:gti}
\end{eqnarray}
where $\varphi$ is the position angle in the counter clockwise direction
from the first coordinate axis on the sky.  
Then, the profiles of $g_{+}$ and $g_\times$ are estimated with a statistical
weight (Equation \ref{eq:ug}), as follows:
\begin{eqnarray}
\langle{g_{\alpha}}\rangle(\theta_n)= \frac{\sum_{i}u_{g,i}
 g_{\alpha,i}}{\sum_i u_{g,i}}.
\label{eq:g}
\end{eqnarray}
Here, $n$ denotes the $n$th radial bin $\theta_n$ with a given bin.
The statistical error of $g_\alpha$ in each radial bin is estimated as
\begin{equation}
\sigma_{g_\alpha}^2(\theta_n) =\frac{1}{2}\frac{\sum_i u_{i}^2 \sigma_{g,i}^2}
{\left(\sum_i u_{i} \right)^2},
\label{eq:sig_g+}
\end{equation}
where the prefactor $1/2$ comes from the fact that $\sigma_{g,i}$ in the rms
is the sum of two distortion components.

We fit the tangential distortion profile with the universal profile proposed by Navarro,
et al. (1996, hereafter NFW profile) and a singular isothermal
sphere (SIS) halo model. 
The NFW halo mass is a prediction of numerical simulations based CDM model. The mass density profiles over a wide range of masses
are well described in the form of 
\begin{equation}
\rho_{\rm NFW}(r)=\frac{\rho_s}{(r/r_s)(1+r/r_s)^2},
\label{eq:rho_nfw}
\end{equation}
where $\rho_s$ is the central density parameter and $r_s$ is the scale
radius. The asymptotic inner and outer slopes for NFW mass density are
$\rho\propto r^{-1}$ and $r^{-3}$, respectively.
The three-dimensional mass profile within a radius, $r_\delta$, at which 
the mean density is
$\Delta$ times the critical mass density, $\rho_{\rm cr}(z)$, at the
cluster redshift, is expressed by
\begin{eqnarray}
M_{\rm NFW}(<r_{\Delta})=
\frac{4\pi\rho_s r_\Delta^3}{c_\Delta^3} m(c_\Delta), \label{eq:Mnfw}
\end{eqnarray}
with
\begin{eqnarray}
m(x)&=&\log(1+x)-\frac{x}{1+x}.
\end{eqnarray}
The NFW halo mass is described by two parameters of the mass $M_{\rm NFW}(<r_\Delta)$ 
and the halo concentration $c_{\rm \Delta}=r_{\Delta}/r_s$.

The SIS halo model is a solution of the collisionless Boltzmann equation, 
and is specified by one parameter, the one-dimensional velocity dispersion
$\sigma_v^2$, as follows,
\begin{equation}
\rho_{\rm SIS}(r)=\frac{\sigma_{v}^2}{2\pi G}\frac{1}{r^2}. 
\label{eq:sis}
\end{equation}
A three-dimensional mass for the SIS model is given by
\begin{eqnarray}
M_{\rm SIS}(<r_{\Delta})&=&\frac{2\sigma_{v}^2}{G} r_\Delta.
\end{eqnarray}

\begin{figure*}
  \begin{center}
   \includegraphics[width=0.55 \textwidth,angle=0,clip]{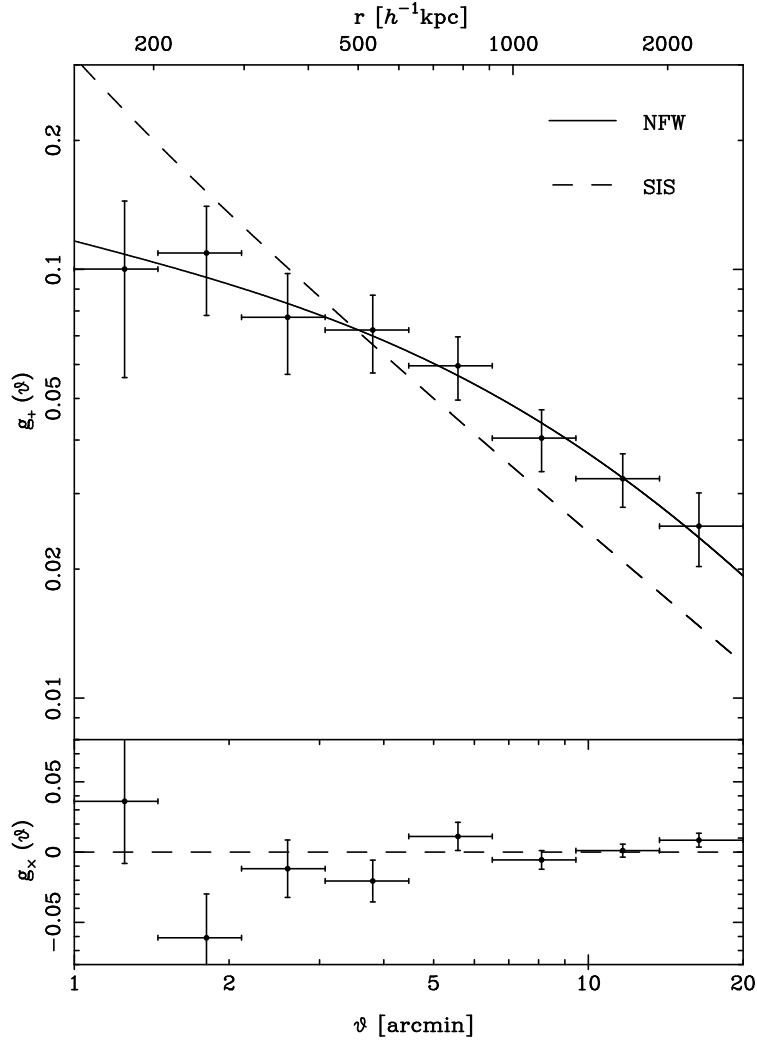} 
  \end{center}
  \caption{Profiles of the tangential shear component (top panel), $g_+$, and the $45$ degree rotated
component (bottom panel), $g_\times$. The solid and dashed lines are the
 best-fit NFW and SIS models, respectively. The SIS model is strongly
 disfavored ($\sim5\sigma$ level) for a cluster mass profile.
}
\label{fig:g+}
\end{figure*}

The tangential distortion profile and the best-fit models are shown in
Figure \ref{fig:g+}. We choose the central position determined by
two-dimensional shear analysis which will be described in detail in
Section \ref{sec:2d}. We also measure the tangential shear
profile with a center of BCG1 and fit them, 
but the results do not change significantly.
We can clearly find a curvature of the tangential shear profile.
The curvature makes it difficult to fit the profile with the SIS
model. The $\chi^2 {\rm (d.o.f)}$ for the SIS model is $27.85(7)$; 
therefore, the SIS model can be strongly rejected ($5\sigma$ level) as a mass model.
On the other hand, the NFW mass model well expresses the curvature of the
profile. Indeed, $\chi^2$ for the NFW model is $0.82(6)$.
The best-fit NFW parameters are found in Table \ref{tab:mass}. The
virial mass shows the massive cluster, $\Mvir =
24.25^{+6.00}_{-4.55}\times10^{14}h^{-1}\Msun$,
where the virial overdensity is $\Delta_{\rm vir}\simeq 116.4$.
Our mass estimates are larger than those of a previous weak-lensing study \citep[Table \ref{tab:mass};][]{rad08}, using the shear catalog
without excluding a contamination of member galaxies.

\section{Two-Dimensional Shear Analysis} \label{sec:2d}

As shown in Figures \ref{fig:opt+kappa} and \ref{fig:kappa+den}, the
projected mass distribution of A2163 is complex and two mass peaks are
significantly detected in the central region. 
It is of prime importance for understanding cluster merger phenomena to
measure masses of the main- and sub- clusters. 
Furthermore, since numerical simulations \citep{men10,bec10} and observations \citep{oka10b}
have shown that a tangential shear profile is affected by significant substructures,
taking into account substructures in modeling is also important for understanding such a lensing bias.
In the mass measurement using the tangential shear profile (Section \ref{sec:1d}),
it is very difficult to 
distinguish which structure contributes in part to the tangential distortion signals,
because the full lensing information from both the main and subclusters must
be convolved to express the one dimensional distortion profile with respect
to a given center \citep{oka10a}.
The two dimensional shear 
pattern, on the other hand, 
enables us to easily model lensing signals by a superposition of lensing
signals. 
In this section, we conduct two-dimensional shear fitting in order to
measure the masses of three components (the sub and main components for A2163-A and A2163-B) 
revealed by M08 and B11.

We pixelize the shear 
pattern into a regular grid of
$1\farcm \times 1\farcm$ without any spatial smoothing procedure, 
whereas we adopted Gaussian smoothing in the map making (Section
\ref{sec:map}). The pixelized distortion signals and statistical weight,
$\langle{g_{\alpha}}\rangle(\btheta_n)$ and
$\sigma_{g}^2(\btheta_n)$,
 in the $n$th pixel are estimated with a weight function $u_i$ for each background source 
residing in the pixel (see also Equations. (\ref{eq:g}) and (\ref{eq:sig_g+})). 
The representative position for the $n$th pixel is also estimated with a
weight function $u_i$.
The $\chi^2$ fitting is given by
\begin{eqnarray}
\chi^2&=&\sum\limits_{\alpha,\beta=1}^{2}\sum\limits_{n}^{N_{\rm pixel}}
 (g_\alpha(\btheta_n)-g_\alpha(\btheta_n;\mbox{\boldmath $p$})^{(\rm
 model)}) C_{\alpha\beta}^{-1}(\btheta_n)  \nonumber \\
&& (g_\beta(\btheta_n)-g_\beta(\btheta_n;\mbox{\boldmath $p$})^{(\rm model)}),
\end{eqnarray}
where $\mbox{\boldmath $p$}$ is the parameters and $C_{\alpha\beta}$ is
the error covariance matrix of shape measurements in the form of
$C_{\alpha\beta}(\btheta_n)=\delta^{\rm
K}_{\alpha\beta}\sigma_g^2(\btheta_n)$. Here, $\delta^{\rm
K}_{\alpha\beta}$ is a Kronecker delta function and
$\sigma_g^2(\btheta_n)$ is the statistical error of the pixelized shear \citep{ogu10,wat11}.


We first consider a single mass model of the NFW profile in order to compare the mass
estimates by tangential shear measurement. We here treat the center of
NFW mass ($x_c$, $y_c$) as a parameter. In total, we use four
parameters ($M$, $c$, $x_c$, and $y_c$) for fitting. 
We adopt the Markov Chain Monte Carlo method with
standard Metropolis-Hastings sampling.
We restrict the sampling range of $M_{\rm
vir}\leq5\times10^{15}h^{-1}M_\odot$, $c_{\rm vir}\le 20$.
We refer to the mean of the posterior probability distribution of each
parameter. 
The resultant masses are consistent with the tangential shear
measurements (Table \ref{tab:mass}).
The central position of mass is consistent with the peak position of the MC
clump in the mass map. 
The MC clump, which is associated with BCG1,
 is therefore likely to be the main component.

We next add a mass model for the mass clump MW to the main cluster. 
From X-ray and optical spectroscopic studies (M08 and B11), the mass
clump MW is likely to be a merging substructure in A2163-A. Since cluster substructure size is not
determined by the virial theorem but by the strong tidal force of the main cluster
\citep[e.g.,][]{tor98}, we adopt the truncated SIS (TSIS) model \citep{oka10a}
to describe the MW clump. The TSIS model is an extreme case of the
truncation; mass density becomes zero at a radius $r_t$.
The TSIS mass profile is expressed as 
\begin{eqnarray}
  \rho_{\rm TSIS}(r) &=& \rho_{\rm SIS}(r) ~~~~~~{\rm for}~~ r \le r_t \label{eq:rhoTSIS}  \\
                  &=& 0      ~~~~~~~~~~~~~{\rm for}~~r > r_t  \nonumber.
\end{eqnarray}
The subclump mass for the TSIS
model is estimated as
\begin{eqnarray}
 M_{\rm sub}^{\rm (TSIS)}= \frac{2\sigma_{v,t}^2}{G}r_t.
\end{eqnarray}
The TSIS model is specified by one-dimensional velocity
dispersion $\sigma_v$ and the truncation radius $r_t$. 
We have an additional four parameters ($\sigma_v$, $r_t$ and centers) for the
TSIS model and a total of eight parameters for the fitting.
We assume that the redshift of the MW clump is the same as that of the main
cluster.  We adopt $|x_c-x_{\rm peak}|<2\farcm$ and $|y_c-y_{\rm peak}|<2\farcm$, where $x_{\rm
peak}$ and $y_{\rm peak}$ are peak coordinates appearing in a weak-lensing mass map. 
The resultant mass and central positions are shown in Table
\ref{tab:mass2}. 
The centroid of the MW clump is consistent with the peak found in the
mass map.  The truncation radius, $r_t=9\farcm7_{-2.9}^{+3.4}\sim 1.4h^{-1}{\rm Mpc}$, is an
intermediate size compared to the virial radius of the main cluster
$r_{\rm vir} = 16\farcm6\sim 2.3h^{-1}{\rm Mpc}$.  Since 
the sub-cluster size after some impacts is significantly decreased by the strong tidal field, 
such a large truncation radius might suggest that the MW clump is a merging
sub-cluster at the first impact.

Next, we take into account the northern component A2163-B (M08 and
B11 ), which contains luminous galaxies (optical clump B) and an X-ray
emitting core. 
We consider two possibilities for the dynamic
state of A2163-B; one scenario is the pre-merger phase that A2163-B is infalling
toward A2163-A, 
and the other is that A2163-B has already undergone
a merging event with the main cluster of A2163-A. 
B11 found no evidence of the close interaction between A2163-A
and A2163-B, and therefore concluded that they are likely to be
separated more than $r_{500}$ aligning along the line of sight. 
However, in M08, a filament of faint galaxies was detected 
along a north/south axis between A2163-A and A2163-B 
which might suggest a previous interaction between the two components. 
Although this post-merger  hypothesis is very unlikely 
from the X-ray approach, the origin of this faint galaxy filamentary 
structure is still an open issue.
In this paper, we investigate two possibilities in the fitting.
If A2163-B is physically separated from A2163-A, its mass profile is not
likely to be affected by the strong tidal field of A2163-A and 
we therefore adopt the NFW model for the first scenario.
Here, the halo concentration is ill constrained
because the shear 
signals from A2163-B are smaller than those from
A2163-A, and it is difficult in the environment of the massive cluster 
to find the curvature of 
the distortion profile, as shown in Figure \ref{fig:g+}. 
We therefore assume the mass - concentration relation of
$\cvir=7.85\left(\Mvir/2\times10^{12}h^{-1}\Msun\right)^{-0.081}(1+z_c)^{-0.71}$
\citep{duf08}. Here, we assume a redshift of A2163-B, $z_c$, is the same as that
of A2163-A. 
This assumption is justified by both the dynamical analysis (M08)
and the photometric redshift estimates by \cite{lab04}.
We also parameterize the center of A2163-B, using 13 parameters in total.
The resultant virial mass is $M_{\rm
vir,B}=2.43_{-1.15}^{+0.90}\times10^{14}h^{-1}M_\odot$ (Table
\ref{tab:mass2}). The mass $M_{500,\rm
B}=1.34_{-0.64}^{+0.49}\times10^{14}h^{-1}M_\odot$ 
within a radius $r_{500}$, at which the mean density is $500$ times 
the critical density, agrees well with X-ray estimates $M_{500,
B}=(1.47\pm0.07)\times10^{14}h^{-1}M_\odot$ (B11).
We changed the normalization of the mass -concentration relation by $\pm2$ 
 and conducted fittings, but the best fit of the virial mass for A2163-B
 changes only by $+14\%$ and $-10\%$. They all are consistent within
 errors. We also found that the MC and MW masses are almost unchanged
 after adding the mass model of A2163-B. 
The resultant mass ratios of MC (main cluster of A2163-A), MW (sub
cluster of A2163-A) and A2163-B are found to be $\sim8:1:1$.

We consider the second scenario that A2163-B is an ongoing merger
with A2163-A. 
We here use the TSIS model for A2163-B, following the case of the substructure MW.
The truncation radius $r_t=741.3^{+289.1}_{-342.8}~{\rm kpc}h^{-1}$ is obtained,
while the concentration of the NFW mass model is not well constrained.
This is because the tangential distortion profile for the truncation model 
outside the truncation radius is proportional to the inverse square of the radius ($g_+\propto r^{-2}$), 
which is different from that of the NFW mass model for the main cluster.
We obtain the substructure mas $M_{\rm B}=1.44_{-0.84}^{+0.64}\times10^{14}h^{-1}M_\odot$. 
The mass discrepancy between the two models is small: 
$0.9\sigma$ and $1.2\sigma$ with uncertainties of NFW and TSIS masses, respectively. 
The two mass models (NFW and TSIS) obtained solely by shear data are, therefore, acceptable for A2163-B.
The total mass in each model is consistent within errors with tangential shear measurement, 
which indicates that a substructure effect is likely to be less significant.

\section{Mass Comparisons} \label{sec:mass}



\begin{figure*}
  \begin{center}
   \includegraphics[width=0.48 \textwidth,angle=0,clip]{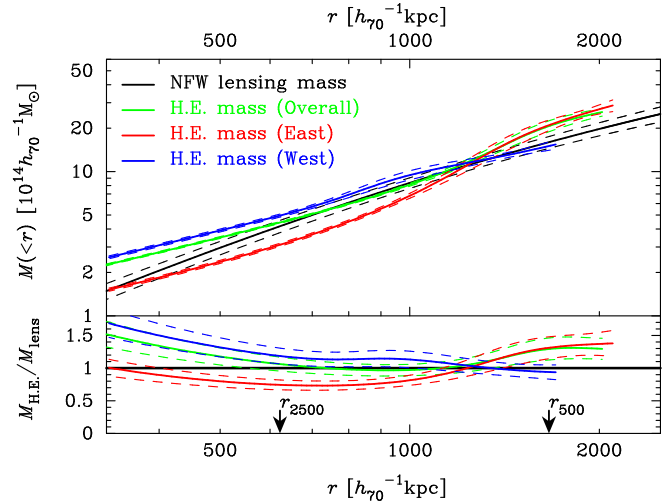} 
  \end{center}
  \caption{Comparison of NFW lensing mass and hydrostatic
 masses. Black color represents the NFW lensing mass profile. Green, red and
 blue colors represent the hydrostatic mass in overall, east and west
 sectors \citep[B11;][]{bou10}, respectively.
The solid and dashed lines are the best fit values and the 68\% CL
 uncertainty errors, respectively.
The arrows denote the overdensity radii $r_{2500}$ and $r_{500}$
 determined by weak-lensing analysis.
We use $h_{70} = H_0 / 70\; \rm{km\; s^{-1}\; Mpc^{-1}}=1.0$.
}
\label{fig:Mhe_vs_Mwl}
\end{figure*}

Due to its exceptionally high mass and dynamical activity, 
A2163 has long been an interesting test case for cluster mass measurements. 
We here compare the weak lensing cluster mass, $\Mvir =
24.24^{+6.00}_{-4.55}\times10^{14}h^{-1}\Msun$ (see Section \ref{sec:1d} and Figure ~\ref{fig:Mhe_vs_Mwl}), 
with dynamical and X-ray mass estimates provided by M08 and B11. 
As previously mentioned, weak lensing mass measurements do not require any assumption of cluster dynamical state. 
They consequently provide us a unique opportunity to investigate effects of the 
virial theorem hypothesis for member galaxies 
or the gas H.E. assumption on mass measurement precision.

\subsection{Mass estimate from galaxy dynamics}

We first compare weak-lensing mass with a dynamical one.
M08 estimates dynamical total mass from the velocity
dispersion, $\sigma_{\rm l.o.s}\simeq1400~{\rm kms}^{-1}$, 
under assumptions of the virial theorem, spherical symmetry, and no
internal structure. 
The virial radius in dynamical mass estimation is defined as the three-dimensional
radius of harmonic radius, corresponding to a coherent length of
galaxy separations, which is different from that of the weak-lensing mass measurement.
We here therefore compare $M_{200}$ enclosed in the radius $r_{200}$ 
within which the mean density is $200$ times the critical density
$\rho_{\rm cr}$. 
Dynamical mass, $M_{200}=(27.3\pm2.8)\times10^{14}M_{\odot }h^{-1}$, 
extrapolated with a Hernquist profile \citep{her90}, is
higher than the weak-lensing mass measured by both one- and two- dimensional
analyses (Table \ref{tab:mass}).
The discrepancy might be due to a difference of models.
We would require a more detailed dynamical model considering 
uncertainty of the anisotropy in the velocity distribution \citep{lok01}.
On the other hand, we calculate one-dimensional velocity dispersion
corresponding to the virial mass for the NFW model
and obtain $\sigma_{\rm 1D}=r_{\rm vir}^{\rm (NFW)} H(z_c)\Delta_{\rm
vir}^{1/2}/2\simeq 1459~{\rm kms}^{-1}$ assuming an isotropic velocity
dispersion, which agrees well with the velocity dispersion along the
line-of-sight (M08). They are higher 
than the one-dimensional velocity dispersion,
$1174.24_{-41.36}^{+39.34}~{\rm kms}^{-1}$, of the SIS model which is
ruled out at the $5\sigma$ level by one-dimensional weak-lensing analysis. 
We note that this is a rough comparison, because the observed
line-of-sight velocity dispersion is calculated with a weight of mass
density \citep{bin82}.

\subsection{X-Ray mass estimates}

\subsubsection{Mass estimate from ICM hydrostatic equilibrium}

We next compare X-ray masses using H.E. assumption.
The impact of the ongoing cluster merger event  in the central region of A2163 
on hydrostatic equilibrium (H.E.) of the hot gas has been investigated in B11. 
To do so, an average cluster mass profile has been extracted from density and 
temperature profiles of the overall cluster atmosphere, in addition to two supplementary
profiles extracted assuming H.E. in the complementary regions located behind and 
ahead the gas core currently crossing A2163 (see also Sect. \ref{sec:mass} and Figure 7 of B11).
These profiles have been compared with our NFW lensing profile in Figure ~\ref{fig:Mhe_vs_Mwl}.
We first find that H.E. mass profiles extracted in the eastern cluster side and overall cluster 
strongly exceed the NFW profile at an overdensity radius of $r_{500}$, and beyond. 
Interestingly, 
we find instead that the H.E. mass profile extracted in the western cluster sector fully agrees with the weak- 
lensing profile in the radii range $r_{2500}$--$r_{500}$. 
Mass estimates corresponding to this analysis at 
the overdensity $\Delta=500$ are reported in Table ~\ref{tab:mass}. 
Adopting overdensity radii determined by each measurement method, 
we observe that the H.E. mass in all and eastern sectors are $\sim1.5$ ($\sim 4\sigma$ level) and $\sim1.7$ times ($\sim 5\sigma$ level) higher than the weak lensing mass, respectively, while the one in the western 
sector is in a good agreement with weak lensing one. 
From this comparison, we may infer that
H.E. assumption is only realistic in the western side of the cluster outer radii.
The eastern cluster side is instead likely to have been shock heated by the ongoing 
subcluster accretion, 
yielding a strong departure of this cluster region and thus of the overall cluster atmosphere from H.E..

\subsubsection{Mass estimate from the $Y_X$ proxy}

One concern in cluster cosmology is the search for X-ray mass proxies relying on well-calibrated scaling
relations coupling the cluster gas properties with total masses \citep[e.g.,][]{kra06, vik09, arn10,zha08,oka10c}. 
\cite{kra06} have proposed a new mass proxy, the so-called, quasi-integrated gas pressure,
$Y_X\equiv M_{\rm gas}T$. This quantity has been suggested by numerical simulations 
to exhibit low intrinsic scatter regardless of the cluster dynamical state, in particular
since deviations of temperature and gas mass from the normalization of 
the mass scaling relations partly anti-correlate during cluster mergers. 
 The cluster mass, $M_{500}$, has been estimated using the $Y_X$ proxy in B11, 
assuming the  $M_{500}$--$Y_X$ scaling relation to be calibrated from hydrostatic 
mass estimates in a nearby cluster sample observed with {\it XMM-Newton} \citep{arn10}.
This mass estimate, $M_{500}=13.09^{+1.40}_{-1.68}\times10^{14}\Msun h^{-1}$, yields
a 20 $\%$ excess with respect to our weak-lensing mass. While marginally significant (1.1$\sigma$ significance level), 
this mass discrepancy slightly exceeds the $\sim6\%$ scatter of the $Y_X$ proxy suggested by 
numerical simulations in \cite{kra06}. \cite{oka10c} have investigated a dynamical
dependence on mass observable scaling relations, based on {\it
XMM-Newton} X-ray observables \citep{zha08} and weak- lensing masses
\citep{oka10b}. This study found the normalization of the $M_{500}$--$Y_X$ scaling relation to be lower for disturbed
clusters than for undisturbed clusters, contradicting the prediction of normalization 
independence in cluster dynamical states. Mass estimates of A2163 relying on the undisturbed and disturbed cluster scaling relation calibrated by \cite{oka10c} yield $M_{500}^{\rm (undist)}\simeq14.53^{+2.5}_{-2.1}\times10^{14}\Msun h^{-1}$ and $M_{500}^{\rm (dist)}\simeq11.4^{+1.8}_{-1.5}\times10^{14}\Msun h^{-1}$, respectively.
Interestingly, the mass estimate with the scaling relation for disturbed clusters fully agrees 
with weak-lensing mass, while the undisturbed cluster estimate yields a higher value, consistent
with the $Y_X$ estimate of B11. Since A2163 is an ongoing merger, the normalization for disturbed 
clusters gives a better result. In order to constrain the cosmological parameters \citep[e.g,][]{vik09,vik09b},
it is of prime importance to construct and calibrate a low-scatter mass proxy.
Recently, Millennium Gas Simulations \citep{sta10} predicted that the temperature and gas-mass deviations
are positively correlated, which contradicts \citet{kra06}.
It is also important to construct a low-scatter mass proxy using solely
observational data, based on intrinsic
covariance measurement and principal component analysis \citep{oka10c}.
Further systematic studies, including both numerical simulation and
observational study, are required for this.

\section{DISCUSSION} \label{sec:dis}

From the inversion of a background galaxy shear pattern,
we revealed a bimodal mass distribution in the central region of A2163 and various
anisotropies at larger radii, the most significant of which
corresponding to the northern subcluster A2163-B. Modeling the
underlying three dimensional distribution of the cluster mass with three components
further allowed us to constrain the mass ratio between the two central
components to 1:8, and attribute comparable mass values to the central
and northern subclusters.

The spatial coincidence between X-ray peak, mass and member galaxy
distribution in A2163-B suggests  that A2163-B did not yet interact
with A2163-A.  While coinciding with the bimodal galaxy distribution
revealed in M08, the central mass distribution appears instead as
spatially segregated from the brightest X-ray emitting gas region. In
particular, the gas bullet suggested to cross the cluster atmosphere
from the evidence of a cold front, in B11, appears as preceded by the
secondary mass peak to a projected distance of $\sim2\farcm\sim280h^{-1}~{\rm kpc}$
along its westward direction of motion. Assuming an infalling subcluster has experienced a
transient separation of its gas and dark matter components as a result
of ram pressure stripping, this configuration would require the gas
core to have survived tidal distortions exerted by the main
cluster, and the core temperature to be consistent with the
subcluster mass. In the following, we discuss how these conditions
allow us to put some additional constraints on the kinematics of the
ongoing subcluster accretion.

\subsubsection*{The Ram Pressure Stripping Condition}
 
One possibility for explaining the offset between mass and gas distributions
is the ram-pressure stripping: 
if the ram-pressure force on the X-ray core is stronger than the gravity
around the core region, the gas is stripped from its gravitational potential.
This condition is expressed as
\begin{eqnarray}
 \frac{G M(< r_{\rm core}) \rho_{\rm core}}{r_{\rm core}^2}
  <  A (\pi r_{\rm core}^2
  \rho_{\rm sur} v^2) \left(\frac{4}{3} \pi r_{\rm core}^3\right)^{-1} \label{eq:Pram}
\end{eqnarray}
\citep{tak06}. Here, $r_{\rm core}$ is the radius of the X-ray core, $M(<
 r_{\rm core})$ is the spherical mass of the subcluster within the radius $r_{\rm
core}$, $\rho_{\rm core,sur}$ are the density of the core and its
surrounding gas, respectively, and $v$ is the velocity of the core in the
 center-of-mass frame.
$A$ is a fudge factor at an order of unity. This is why the ram-pressure
stripping is not effective due to a Kelvin-Helmholtz instability,
magnetic fields, and shock heating.
B11 shows a lower limit of the density jump, $\rho_{\rm core}/\rho_{\rm sur}=1.28$, 
at the west edge of the core (red-curved line in Figure \ref{fig:chan+kappa}).

\begin{figure*}
  \begin{center}
   \includegraphics[width=0.55 \textwidth,angle=0,clip]{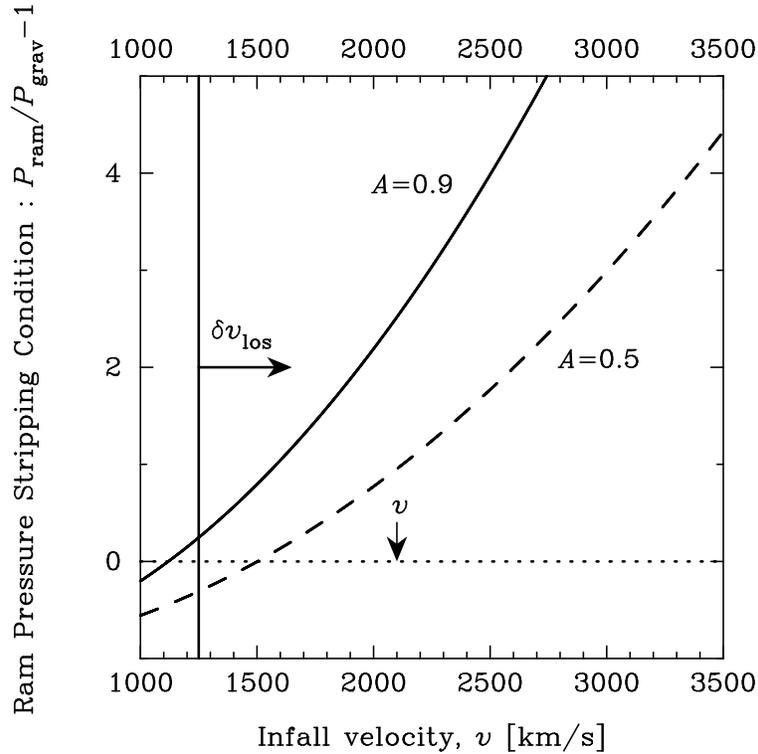} 
  \end{center}
  \caption{Ram pressure stripping condition of the subcluster core.
  $A$ is a fudge factor associated with hydrodynamical
 instabilities or gas heating. The vertical line denotes the
 gradient of the line-of-sight velocity (M08). The arrow denotes
 an order of the infall velocity (Equation (\ref{eq:vmerger})) expected from masses for main and sub clusters,
 measured by two-dimensional weak-lensing analysis.
}
\label{fig:Pram}
\end{figure*}

We consider the ram-pressure stripping condition on the sub cluster. 
The substructure mass inside $r_{\rm core}$ is calculated from the TSIS
model obtained by the weak-lensing analysis.  Since $M_{\rm TSIS} \propto
r$, the condition (\ref{eq:Pram}) is free from the core size. 
Although we consider the internal structure of gas, the resultant
condition does not significantly change.
Figure \ref{fig:Pram} plots the ram-pressure stripping condition $P_{\rm
ram}/P_{\rm grav}-1$ as a function of infall velocity $v$,  
where $P_{\rm ram}/P_{\rm grav}$ gives the ratio of the right-hand side to the
left-hand side of Equation (\ref{eq:Pram}). 
We adopt $A=0.5~{\rm and}~0.9$.
M08 found the gradient of line-of-sight velocity $v_{\rm
los}\sim1250~{\rm km~s^{-1}}$, giving the lower-limit of infall velocity.
The infall velocity is expected from the masses to have an order of 
\begin{eqnarray}
v\sim \left[\frac{2G(M_{\rm main}+M_{\rm sub})}{r_{\rm main}+r_{\rm
       sub}}\right]^{1/2}\sim 2100 ~{\rm km~s^{-1}} \label{eq:vmerger}
\end{eqnarray}
\citep{ric01}, where $M$ and $r$ are the virial mass and radius, respectively.
We here used observed truncated mass and radius for the sub cluster.
The case of $A=0.9$ satisfies the ram-pressure stripping condition for
all velocities above $\delta v_{\rm los}$.
The case of $A=0.5$ also satisfies the condition around $2100~{\rm
km~s^{-1}}$. Therefore, the  X-ray core initially associated with the sub cluster
is easily stripped away from its central region.
However, we must keep in mind that the estimated density jump might give the lower limit, 
due to the deprojection method we applied.
In this case, a requirement of the ram pressure condition gives 
$\rho_{\rm core}/\rho_{\rm sur}<5(A/1)(v/2100~{\rm km~s^{-1}})^2$.


\subsubsection*{Constraints on impact parameter from survival condition}

We next discuss the destruction process of the gas core by the strong tidal field of
the main cluster.
Numerical simulation \citep{tor04} has shown that the lifetime for a gas
satellite is shorter than that for a dark matter one. 
In particular, the lifetime for a gas core decoupled from dark matter
potential is much shorter than that of a dark matter satellite. 

\begin{figure*}
  \begin{center}
   \includegraphics[width=0.8 \textwidth,angle=0,clip]{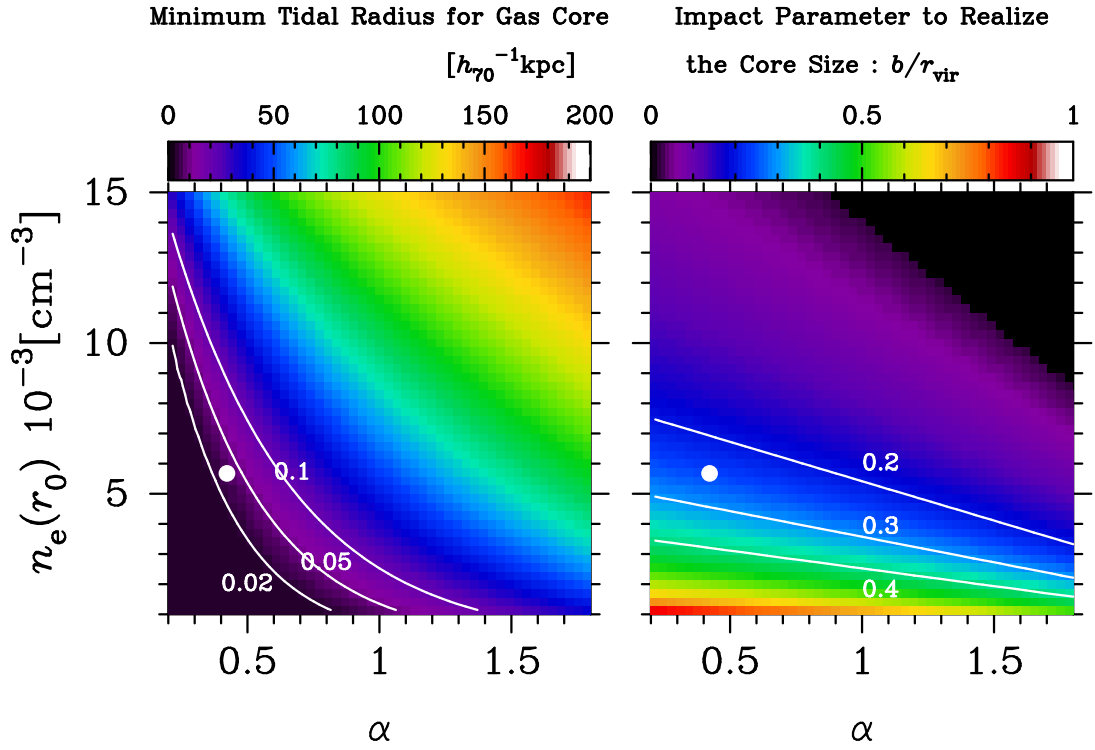} 
  \end{center}
  \caption{Left : minimum tidal radius in the parameter plane of
 core density normalization, $n_e(r_0)=n_{e,0}$ at the cold front $r=r_0$
and slope $\alpha$. The
 contours represent tidally truncated core sizes of $0\farcm02,
 0\farcm05$, and $0\farcm1$ in units of arcmin.  
The point denotes the observed parameters for the core XW.
Right: impact parameter to realize the observed core size. The
 contours are $(0.2-0.4)r_{\rm vir}$. The point is the same as that in the
 left panel. 
}
\label{fig:taidal}
\end{figure*}

We roughly estimate the gas core disruption by the tidal force of the main cluster.
As shown above, the gas core could easily be stripped away from the
subcluster's core region. 
We therefore consider 
the stripped core which is composed of the gas, to be free from the gravity of the subcluster.
The gas model for the core is adopted as a single power-law density profile as an approximation,
\begin{eqnarray}
  n_e(r)=n_{e,0} \left(\frac{r}{r_0}\right)^{-\alpha},
\end{eqnarray}
where the normalization $n_{e,0}$ and the slope $\alpha$ are the model
parameters. We use the position of cold front $r_0=0\farcm6$ 
(red-curved line of the right panel of Figure \ref{fig:chan+kappa}).
The tidal radius at which the core density is truncated is estimated
by the force equivalence between the internal gravity and external tides
\citep{tor98},  
\begin{eqnarray}
 r_t=\tilde{A} b\left[\frac{M_{\rm core}(<r_t)}{(2-\partial \ln M_{\rm
       main}/\partial \ln b) M_{\rm main}(b)}\right]^{1/3}, \label{eq:tidal}
\end{eqnarray}
in the limit of $r_t\ll b$ and $M_{\rm core}\ll M_{\rm main}$.
Here $r_t$ is the tidal radius and $b$ is the core position from the
center of the main cluster; that is, an impact parameter.
$\tilde{A}$ is a fudge factor because hydrodynamic instabilities effectively destruct the gas core.
We assume $\tilde{A}=1$ in order to consider the tidal disruption only.
We also assume that the density profile within a tidal radius 
dose not modify before and after the tidal stripping. 
The main cluster mass profile uses the best-fit NFW model in the case of NFW+TSIS+NFW (Table \ref{tab:mass2}).
We first calculate the minimum tidal radius $r_{t,{\rm min}}$ 
as a function of the density normalization and its slope and impact parameter. 
The left panel of Figure \ref{fig:taidal} shows a distribution of the minimum tidal radius of the gas core 
in the parameter plane of the density normalization and the slope.
The contours denote the minimum tidal radius in the range of $0\farcm02-0\farcm1$.
The point denotes the normalization and slope parameters for the core XW at the cold front.
In the parameter plane, the minimum tidal radius is much smaller than the
observed radius $r\sim0\farcm6\sim120h_{70}^{-1}~{\rm kpc}$, indicating 
that we cannot observe a remnant of the gas core if A2163-A is a system of close encounter.

We next try to constrain the impact parameter, $b$, 
so as to realize the observed radial size of the core XW ($r_t=r_0$) through Equation. (\ref{eq:tidal}).
The right panel shows a distribution of the impact parameter 
in the parameter plane of the density normalization and the slope.
The contours present the impact parameters $(0.2-0.4)r_{\rm vir}$ to explain the core size.
The observed X-ray core requires the impact parameter $b\sim0.25 r_{\rm
vir}\sim560~{\rm kpc}h^{-1}$. 
If a subcluster collides into the main cluster with this
impact parameter, the central region of the main cluster is significantly
affected by cluster merger. It does not conflict with the disturbed core
structures and the complex temperature distribution (Figure \ref{fig:chan+kappa}).
Since this estimation is based on the gravitational process only, 
we need to keep in mind a possibility that hydrodynamic instabilities shorten the lifetime of gas.
We emphasize that this method demonstrates only one of the methods for joint X-ray and weak-lensing analysis.





 

\subsubsection*{Temperature Comparison}

Assuming the cool gas core XW and the secondary mass peak MW  
represent the separated components of a formerly accreted subcluster,
the intrinsic temperature of XW should agree with the expectation of a
self-similar subcluster temperature corresponding to the mass of
MW. Given our multi-component mass distribution (NFW+TSIS+NFW; see
Table 2), we discuss the consistency of this hypothesis with projected
ICM temperatures derived near the cool gas core XW in B11.

The cool gas core XW being located a relatively close projected
distance from the main cluster emission peak XC, average ICM temperatures 
measured along a line of sight intercepting XW would consist of a linear 
combination of the main and subcluster temperatures. These temperatures can be
predicted within the radial range [0.2-0.5]$~r_{\mathrm{500}}$, from the
$M_{500}-T_{[0.2-0.5]~r_{500}}$ scaling relation
calibrated by \cite{oka10c}, in a sample of 12 local clusters 
analyzed in both X-ray and weak-lensing. To perform this
estimation, we first measure the main cluster mass, 
$M_{500,{\rm main}}=7.59_{-1.57}^{+1.22}\times10^{14}h^{-1}\Msun$, 
by multi-component analysis (NFW+TSIS+NFW), 
within the overdensity radius $r_{500}$. 
We further determine the subcluster mass, $M_{500,{\rm sub}}$,
from its current 
mass, $M_{\rm sub} = M_{\rm main}/8$, 
assuming the subcluster mass profile to have followed a
universal NFW radial distribution before the ongoing accretion, 
and the profile concentration to be related to $M_{sub}$ 
by the halo mass-concentration dependence of \cite{duf08}.
These assumptions would yield a cluster and subcluster temperature prior to the
accretion of $k_BT = 10.1^{+1.0+1.1+1.8}_{-1.4-1.0-1.8}~\mathrm{keV}$ and 
$k_BT = 3.7^{+0.5+0.3+1.2}_{-0.9-0.3-1.2}~\mathrm{keV}$, respectively, where the first, 
second and third errors are from mass measurement error,
normalization error of scaling relation, and intrinsic scatter in the
relation. Assuming the main and subcluster emissivities to be
comparable from a rough analysis of the X-ray image and weighting these 
temperatures following a scheme proposed in \citet{maz04}, an average
 ICM temperature measured in the direction of the cool core would reach 
 a value of $k_BT \sim 5.7~\mathrm{keV}$. Despite being
consistent with uncertainties in our mass estimates, this value is
considerably lower than projected temperatures measured in 
the cool gas core XW in the \textit{XMM-Newton} temperature map and the \textit{Chandra} 
temperature profile of B11 ($k_BT \simeq 9\pm1~\mathrm{keV}$). 
Among the possible origin for this discrepancy, 
the cool gas core might have been heated 
from its virial temperature while crossing the main cluster
atmosphere, possibly due to mixing with the main cluster atmosphere
or a reverse shock, presumably at an earlier stage of its accretion 
and prior to the formation of the cold front.

\subsubsection*{}

In summary, the density of the main cluster atmosphere and the
velocity of its presumably free-falling subcluster are large 
enough to have separated the subcluster gas and dark matter components 
through ram-pressure stripping. Moreover, the survival of the subcluster gas 
core against tidal forces exerted by the main cluster implies the subcluster accretion to have 
occurred with a non-zero impact parameter. Comparing the average X-ray temperature in the direction of
the cool core to its expectation from projection of the cluster and
subcluster virial temperatures yields however a mild inconsistency,
suggesting the cool core was partially heated while crossing
the main cluster atmosphere.
 Deeper X-ray observations should help better constrain the shape, temperature and mass of the cool core
and refine this accretion scenario.

\section{Summary} \label{sec:sum}

We presented the weak-lensing analysis of the merging cluster A2163 using the
 Subaru/Suprime-Cam and CFHT/Mega-Cam data, measured cluster mass by one-dimensional tangential shear analysis, 
and measured three components' masses by multi-component analysis of the two-dimensional shear pattern.
Based on complementary X-ray, dynamical and weak-lensing datasets, 
we also discussed the centroid offset between weak-lensing mass and gas core.
 Our main results are summarized below.

\begin{itemize}


\item The projected mass distribution shows a bimodal structure in
      the central part of A2163-A. 
      The overall mass distribution appears to be similar to the
      member galaxy one,
      whereas both mass and member galaxy distributions are completely different
      from the ICM one.
      This is consistent with a previous study of seven
      merging clusters at various dynamical state \citep{oka08}.
      In particular, we found a clear offset between the gas core
      associated with the cold front and sub-cluster mass peak.
      The offset was reported in all cold-front clusters previously conducted
      by weak-lensing analysis \citep{clo06,oka08,oka10b}. 
      The gas core is also offset from the main cluster mass peak,
      like the bullet cluster \citep{clo06}.


\item  A two dimensional shear analysis has enabled us to measure the mass of each of
	the three major components in A2163, including the two components of A2163-A, and the northern 
	subcluster A2163-B. The central subcluster in A2163-A is well described by the TSIS model,
        giving the mass
        $2.08^{+0.96}_{-0.97}\times10^{14}h^{-1}\Msun\sim M_{\rm main}/8$.

\item   The mass of A2163-B is comparable to the
        sub cluster of A2163-A, which is in good agreement with the X-ray
        estimate of B11, assuming H.E. 
        The mass, member galaxies, 
        and gas distributions are similar to one another, as reported in pre-merging clusters
         \cite{oka08}, suggesting that A2163-B did not yet interacted with A2163-A.


\item   The gas bullet suggested to cross the cluster atmosphere
	from the evidence of a cold front, in B11, appears to be preceded by the
        secondary mass peak to a projected distance of $\sim2\farcm\sim280h^{-1}~{\rm kpc}$. 
        We show that the density of the main cluster 
      	atmosphere and the free fall velocity of an incoming subcluster 
        with which mass is about one-eighth of main cluster virial mass
        are large enough to have 
	separated the dark matter and  gas component of this subcluster through
	ram-pressure stripping. 
        Following this scenario, the survival of the cool core 
	against tidal forces exerted by the main cluster lets us infer that the subcluster must have been accreted 
	with a non-zero impact parameter, reaching typical values of $b\sim0.25 r_{\rm vir}\sim560~{\rm kpc}h^{-1}$. 
	Assuming the cool gas core and the secondary mass peak represent the separated components of the formerly 
	accreted subcluster, the projected core temperature appears higher than expected from our estimates 
	of the main and subcluster virial temperatures. Subsequent to its separation with dark matter, the crossing 
	cool core may thus have been shock heated or disturbed by hydrodynamical instabilities 
	yielding a partial mixing with the main cluster atmosphere.

\item   Dominated by the most massive component of A2163-A, the overall
mass distribution in A2163 is well described by a universal NFW profile as shown by a tangential distortion analysis,
while the SIS profile is strongly rejected ($5\sigma$
confidence level). The virial mass for the NFW mass model is $M_{\rm
       vir}=24.23^{+6.00}_{-4.55}\times10^{14}h^{-1}\Msun$.

\item We compare the weak-lensing NFW mass with dynamical and X-ray H.E. ones. 
      The weak-lensing mass is lower than the dynamical one with
      assumptions of the virial theorem and spherical symmetry mass
      distribution. This might be due to the dynamical mass model, because 
      the line-of-sight velocity dispersion expected from the NFW mass agrees well with 
      the spectroscopic result (M08).
      The H.E. mass, $M_{500}$, in the western sector showing no hot gas,
      is in good agreement with the weak-lensing one,
      whereas the one in the eastern sector showing hot gas is higher at the $5\sigma$ level. 
      This indicates that the merger
      shock heating leads to an overestimation of H.E. mass. The mass proxy,
      $Y_X$, through the mass observable scaling relation for disturbed
      clusters \citep{oka10c} gives a mass estimate similar to
      the weak-lensing mass. 



\end{itemize}

\section*{Acknowledgments}

NO acknowledges Yoichi Oyama, Keiichi Umetsu, Motoki Kino, Keiichi Asada, Makoto
Inoue and Alberto Cappi for helpful discussions. 
We thank the anonymous referee for the useful comments that led to improvement of the manuscript.
We also thank G. Soucail for giving us CFHT $g'$-band data.
We are also grateful to N. Kaiser for developing the publicly available IMCAT
package. H.B and P.M acknowledge support by NASA grants NNX09AP45G 
and  NNX09AP36G grant ASI-INAF I/088/06/0 and ASI-INAF I/009/10/0.

\clearpage

\begin{table}
  \caption{Optical subclumps} \label{tab:opticalclumps}
\begin{center}
\begin{tabular}{l||c|c}
\hline
\hline
Names &  $L_{\rm R_c}(<2\farcm)$ & S/N \\
(1)   & (2)     & (3)  \\
\hline
C     & $1.47$ & $2.0$ \\
D     & $0.94$ & $1.7$ \\
E     & $2.38$ & $2.7$\\
\hline
\end{tabular}
\end{center}
\tablecomments{Column (1): names of optical subclumps identified by M08.
  Column (2): the luminosity within $2\farcm$ in units of $10^{11}L_{\sun}h^{-2}$.
  Column (3): signal-to-noise ratios appearing in the mass map (Figures \ref{fig:opt+kappa} and \ref{fig:kappa+den}). }
\end{table}

\begin{table}
  \caption{Best-fit Mass Profile Parameters} \label{tab:mass}
\begin{center}
\begin{tabular}{l||cc|cc|cc}
\hline
\hline
Method  & $\Mvir$
        & $\cvir$ 
        & $M_{200}$
        & $c_{200}$ 
        & $M_{500}$
        & $c_{500}$ \\
        & $(10^{14}h^{-1}M_\odot)$
        &
        & $(10^{14}h^{-1}M_\odot)$
        &
        & $(10^{14}h^{-1}M_\odot)$
        & \\
(1)     & (2)
        & (3)
        & (4)
        & (5)
        & (6)
        & (7)\\
\hline
1D WL  &  $24.24_{-4.55}^{+6.00}$
    &  $2.84_{-0.61}^{+0.71}$ 
    & $18.86_{-3.17}^{+3.99}$
    & $2.18_{-0.50}^{+0.57}$
    &  $11.18_{-1.46}^{+1.64}$
    &  $1.35_{-0.33}^{+0.39}$ \\
2D WL &  $26.72_{-5.79}^{+4.72}$ 
    &  $2.58_{-0.59}^{+0.48}$
    & $20.32_{-4.08}^{+3.21}$
    & $1.99_{-0.49}^{+0.39}$
    & $11.13_{-1.55}^{+1.31}$
    & $1.37_{-0.30}^{+0.20}$\\
Dynamics (M08)  & --
                    & --
                    & $27.3\pm2.8$
                    & --
                    & --
                    & -- \\
X-ray (B11) : H.E.                       & --
                           & --
                           & --
                           & --
                           & $17.29^{+0.35}_{-0.63}$
                           & -- \\
X-ray : H.E. west  & --
           & --
           & --
           & --
           & $10.15^{+0.42}_{+0.21}$
           & --\\
X-ray : H.E. east  & --
           & --
           & --
           & --
           & $18.55^{+1.33}_{-0.84}$
           & -- \\
WL (Radovich et al. 2008) &  $15.4\pm2.8$
                          &  --
                          &  $12.6\pm2.1$
                          &  -- 
                          &  $8.4\pm1.4$
                          &  -- \\
\hline
\end{tabular}
\end{center}
\tablecomments{ The best-fit NFW parameters obtained by 
fitting to the one-dimensional tangential distortion profile (1D) and
 the two-dimensional shear 
pattern (2D). For comparison, the masses
 derived from dynamical \citep[][M08]{mau08}, X-ray \citep[][B11]{bou10}, and weak-lensing analyses \citep{rad08} 
are quoted. 
H.E., H.E. west and H.E. east are hydrostatic masses for all, western, and eastern 
sectors, respectively. X-ray mass estimations uses $r_{500}$ determined solely by X-ray analysis.
Weak-lensing masses are obtained using the shear catalog without excluding a contamination of member galaxies.
   Column (1): method.
  Column (2) and (3): virial mass in units of  $10^{14}h^{-1}M_\odot$ and
 the concentration parameter for the NFW model.
  Column (4) and (5):  $M_{200}$ for the overdensity of $\Delta=200$ and
 concentration parameter.
 Column (6) and (7):  $M_{500}$ for the overdensity of $\Delta=500$ and
 concentration parameter.
}
\end{table}

\begin{table}
  \caption{Multi-component Analysis of the Two-dimensional Shear Pattern} \label{tab:mass2}
\begin{center}
\begin{tabular}{l||ccc}
\hline
\hline
Parameters & NFW+TSIS & NFW+TSIS+NFW & NFW+TSIS+TSIS \\
(1)        & (2)        & (3)        & (4) \\
\hline
$M_{\rm vir,main}$ & \scriptsize{$20.98_{-7.38}^{+4.67}\times10^{14}h^{-1}M_\odot$}
                   & \scriptsize{$16.46_{-5.90}^{+3.69}\times10^{14}h^{-1}M_\odot$}
                   & \scriptsize{$18.79_{-6.68}^{+4.31}\times10^{14}h^{-1}M_\odot$}\\
$c_{\rm vir,main}$ & \scriptsize{ $2.88_{-0.82}^{+0.70}$}
                   & \scriptsize{$3.44_{-1.05}^{+0.78}$}
                   & \scriptsize{$3.14_{-0.96}^{+0.74}$} \\
$M_{\rm MW}$  &\scriptsize{  $2.13_{-0.85}^{+0.95}\times10^{14}h^{-1}M_\odot$} 
              & \scriptsize{$2.08_{-0.97}^{+0.96}\times10^{14}h^{-1}M_\odot$}
              &\scriptsize{ $2.09_{-0.96}^{+0.96}\times10^{14}h^{-1}M_\odot$} \\
$r_{t,{\rm MW}}$ & \scriptsize{$1.39_{-0.38}^{+0.47}h^{-1}{\rm Mpc}$}
               & \scriptsize{$1.35_{-0.41}^{+0.49}h^{-1}{\rm Mpc}$}
               &\scriptsize{ $1.37_{-0.39}^{+0.48}h^{-1}{\rm Mpc}$}\\
$M_{\rm B}$  &  --
             & \scriptsize{$2.43_{-1.15}^{+0.90}\times10^{14}h^{-1}M_\odot$}
             & \scriptsize{$1.44_{-0.84}^{+0.64}\times10^{14}h^{-1}M_\odot$} \\
($\alpha$, $\delta$)$_{\rm{c,main}}$ 
              & \scriptsize{ ($243.958_{-0.002}^{+0.003}$, $-6.147_{-0.003}^{+0.003}$)}
              & \scriptsize{ ($243.958_{-0.003}^{+0.003}$, $-6.146_{-0.003}^{+0.003}$)}
              & \scriptsize{  ($243.958_{-0.003}^{+0.004}$,  $-6.147_{-0.003}^{+0.003}$)} \\
($\alpha$, $\delta$)$_{\rm{c,MW}}$ 
              &\scriptsize{($243.905_{-0.005}^{+0.008}$, $-6.125_{-0.005}^{+0.007}$)}
              & \scriptsize{($243.905_{-0.005}^{+0.009}$, $-6.125_{-0.005}^{+0.007}$) }
              &\scriptsize{ ($243.905_{-0.005}^{+0.009}$,  $-6.125_{-0.005}^{+0.008}$)} \\
($\alpha$, $\delta$)$_{\rm{c,B}}$
              & --
              & \scriptsize{($243.969_{-0.005}^{+0.009}$, $-6.042_{-0.005}^{+0.005}$) }
              & \scriptsize{($243.968_{-0.005}^{+0.007}$, $-6.042_{-0.006}^{+0.003}$) }\\
\hline
\end{tabular}
\end{center}
\tablecomments{  The best-fit parameters obtained by fitting
the two-dimensional shear 
pattern with a multi-component model of
 clusters: NFW parameters ($M_{\rm vir,main}$, $c_{\rm vir,main}$) for
the main cluster (MC) at the virial over-density,
  the truncated-SIS (TSIS) mass ($M_{\rm MW}$) for the subcluster (MW), 
 its truncation radius ($r_{t,{\rm MW}}$) in units of ${\rm Mpc}h^{-1}$, and the mass of A2163-B ($M_{\rm B}$). 
The best-fit halo centers are in units of degree.
  Column (1): parameters.
  Column (2): NFW and TSIS models for the main cluster (MC) and for the
 substructure (MW), respectively. 
  Column (3):  NFW, TSIS and NFW models for the main cluster (MC), the subcluster (MW) and A2163-B, respectively. The mass $M_{\rm
 B}$ is the virial mass.
  Column (4):  the NFW model for the main cluster (MC), the TSIS model for
 the subcluster (MW) and A2163-B, respectively. $M_{\rm
 B}$ is the mass within the truncation radius.
}
\end{table}




\end{document}